\documentclass[conference]{IEEEtran}
\IEEEoverridecommandlockouts
\usepackage{algorithmic}
\usepackage[linesnumbered,ruled,vlined]{algorithm2e}
\usepackage{multirow}
\usepackage{multicol}
\usepackage{wrapfig}
\usepackage{soul}
\usepackage{balance}
\usepackage{epstopdf}

\usepackage{amsfonts}
\usepackage[flushleft]{threeparttable}
\usepackage{adjustbox}
\usepackage{fancyhdr}
\usepackage[dvipsnames]{xcolor}
\usepackage[flushleft]{threeparttable}
\usepackage{tikz}
\usepackage{outlines}
\usepackage{tcolorbox}
\usepackage{blindtext}
\usepackage{xcolor}
\usepackage{color}
\usepackage{listings}
\usepackage{color}
\usepackage{MnSymbol,wasysym}
\usepackage[switch]{lineno}
\usepackage{array}

\usepackage{amsmath,amsthm}

\usepackage{mathtools} % Bonus

\usepackage{subfigure}
\usepackage{subfig}
\usepackage[mathscr]{euscript}

\usepackage{caption}
\usepackage{subcaption}

\usepackage{xcolor,colortbl}
\def\BibTeX{{\rm B\kern-.05em{\sc i\kern-.025em b}\kern-.08em
    T\kern-.1667em\lower.7ex\hbox{E}\kern-.125emX}}

\usepackage{tikz}
\newcommand*\circled[1]{\tikz[baseline=(char.base)]{
            \node[shape=circle,draw,inner sep=2pt] (char) {#1};}}

\usepackage[T1]{fontenc} % unrelated to your question, but I'd recommend it

\theoremstyle{plain}

\newcommand\blfootnote[1]{%
  \begingroup
  \renewcommand\thefootnote{}\footnote{#1}%
  \addtocounter{footnote}{-1}%
  \endgroup
}

\begin{document}

% \title{Sparse MTTKRP Acceleration for \\ Tensor Decomposition: A Simple but Elegant Hardware-Algorithm Fusion on GPU}
\title{Accelerating Sparse MTTKRP for \\ Small Tensor Decomposition on GPU}

\author{
    \IEEEauthorblockN{Sasindu Wijeratne\IEEEauthorrefmark{1}, Rajgopal Kannan\IEEEauthorrefmark{2}, Viktor Prasanna\IEEEauthorrefmark{1}}
    \IEEEauthorblockA{\IEEEauthorrefmark{1} University of Southern California, Los Angeles, USA}
    \IEEEauthorblockA{\IEEEauthorrefmark{2} DEVCOM ARL Army Research Office, Los Angeles, USA}
    Email: \{kangaram, prasanna\}@usc.edu, rajgopal.kannan.civ@army.mil
}

%\title{Tensor Formats, Dynamic Tensor Remapping, and Load Balancing: Accelerating spMTTKRP for Tensor Decomposition on GPU}

% \and
% \and
% \and
% \and
% \and

\maketitle

\begin{abstract}
Sparse Matricized Tensor Times Khatri-Rao Product (spMTTKRP) is the bottleneck kernel of sparse tensor decomposition. In tensor decomposition, spMTTKRP is performed iteratively along all the modes of an input tensor. In this work, we propose a mode-specific tensor layout on GPU that uses multiple tensor copies, where each copy is optimized for a specific mode. The proposed tensor layout increases the data locality of external memory accesses and eliminates the intermediate values communicated between the GPU thread blocks and the GPU global memory. We also propose a tensor partitioning scheme to optimally distribute the total computations among GPU streaming multiprocessors based on the sparsity and the dimensions of the input tensor. Our approach achieves a geometric mean speedup of 2.4$\times$, 7.9$\times$, and 8.9$\times$ in total execution time compared with the state-of-the-art GPU baselines.
\end{abstract}


% \vspace{-3mm}
\section{Introduction}
\blfootnote{\textbf{Distribution Statement A:} Approved for public release. Distribution is unlimited.}
Recent advances in collecting and analyzing large datasets have led to the information being naturally represented as higher-order tensors. Tensor Decomposition transforms tensors to a reduced latent space, which can then be leveraged to learn salient features of the underlying data distribution.

This paper focuses on small tensors, specifically those for which all tensor copies corresponding to the matricizations of the tensor can fit in the GPU global memory. Decomposing small tensors (i.e., small tensor decomposition) is used in various domains such as natural sciences~\cite{6811401, yahyanejad2019survey}, medicine~\cite{10.1162/dint_a_00117}, machine learning~\cite{8884203}, networking~\cite{wang2017identifying}, and signal processing~\cite{7891546}. Canonical Polyadic Decomposition (CPD) is arguably the most common means of tensor decomposition. In CPD, the Matricized Tensor Times Khatri-Rao Product (MTTKRP) is iteratively computed along all matricizations of the input tensor, offering a robust methodology for uncovering latent structures within complex data tensors.

CPD approximates a tensor as a sum of rank-one tensors~\cite{doi:10.1137/18M1203626}. Since real-world tensors are generally sparse, keeping only nonzero values in memory is a natural way of reducing the tensor memory footprint. Even though modern GPUs have enough global memory space to store multiple copies of small tensors, there are inherent challenges in accelerating MTTKRP on GPUs, such as irregular memory access patterns~\cite{9622851}, load imbalance between GPU threads and streaming multiprocessors (SMs), and synchronization overhead among SMs.

In this paper, we propose a mode-specific tensor format where each matricization of the input tensor has a unique tensor copy stored in the GPU global memory. Our proposed tensor format eliminates the communication of intermediate values to the GPU global external memory during the execution time of spMTTKRP along all the modes of the input tensor. We also introduce GPU-specific optimizations and balance the total workload among GPU SMs.

The key contributions of this work are:

\begin{itemize}
\item We introduce a novel mode-specific tensor format that eliminates the communication of intermediate values to the GPU global memory.

\item We introduce a novel parallel algorithm to perform spMTTKRP on GPU. The proposed algorithm enables optimal load balancing among streaming multiprocessors and achieves a 1.3$\times$ - 2.2$\times$ speedup in total execution time compared to the conventional load balancing schemes in the literature.

\item Our approach achieves a geometric mean speedup of 2.4$\times$, 7.9$\times$, and 8.9$\times$ in total execution time compared with the state-of-the-art GPU baselines.
\end{itemize}

% The remainder of this paper is structured as follows: Section~\ref{background} focuses on the background and prior work. Section~\ref{Data_partitioning} provides an in-depth exploration of the tensor format, while Section~\ref{subsecparallel_algo} introduces the proposed parallel algorithm. Section~\ref{experiments} presents the experimental results. Finally, our conclusion is summarized in Section~\ref{conclusion}.

\section{Background and Related Work}\label{background}

\subsection{Introduction to Tensors}\label{background:intro}\label{background:decomp}

\begin{wraptable}{r}{0.32\textwidth}
% \begin{table}[ht]
\vspace{-5mm}
\caption{Notations}
\vspace{-5mm}
\begin{center}
\begin{tabular}{|c|c|}
\hline
     \textbf{Symbol} & \textbf{Details} \\
     \hline
          $\circ$ & Vector outer product \\
     $\otimes$ & Kronecker product \\
     $\odot$ & Khatri-Rao product \\
$\mathbf{A}$ & Matrix \\
$\mathbf{a}$ & Vector \\
     $a$ & Scalar \\
     $\mathcal{X}$ & Sparse tensor \\
     $\mathcal{X}_{(d)}$ & Mode-$d$ matricization of $\mathcal{X}$ \\
     \hline
\end{tabular}
\label{table:notation}
\end{center}
\vspace{-5mm}
% \end{table}
\end{wraptable}

A tensor is a generalization of an array to multiple dimensions. In the simplest high-dimensional case, a tensor is a three-dimensional array, which can be visualized as a data cube. For a thorough review of tensors, refer to~\cite{kolda2009tensor}. Table~\ref{table:notation} summarizes the tensor notations.

% In this Section, we will refer to the 3-dimensional tensor for simplicity.

\subsubsection{Tensor mode} In Tensor Decomposition, the number of dimensions of an input tensor is commonly called the number of tensor modes. For example, a vector can be seen as a mode-1 tensor. A $N$-mode, real-valued tensor is denoted by $\mathcal{X} \in \mathbb{R}^{I_0 \times \cdots \times I_{N-1}}$. In this paper, we focus on tensors with three modes or higher.

\subsubsection{Indices of a nonzero tensor element}~\label{index_intro}
For a 3-mode tensor, $\mathcal{X} \in \mathbb{R}^{I_0 \times I_1 \times I_2}$, a nonzero tensor element is indicated as $x = \mathcal{X}(i_0,i_1,i_2)$. Here, $i_0$, $i_1$, and $i_2$ are the positions or coordinates of $x$ in the tensor $\mathcal{X}$, commonly referred to as indices of the tensor element.

\subsubsection{Tensor matricization} $\mathcal{X}_{(n)}$ denotes the mode-$n$ matricization or matrix unfolding~\cite{favier2014overview} of $\mathcal{X}$. $\mathcal{X}_{(n)}$ is defined as the matrix $\mathcal{X}_{(n)} \in \mathbb{R}^{I_n \times (I_0 \cdots I_{n-1} I_{n+1} \cdots I_{N-1})}$ where the parenthetical ordering indicates, the mode-$n$ column vectors are arranged by sweeping all the other mode indices through their ranges.

\subsubsection{Small Tensors} In this paper, we define small tensors as the tensors whose all the tensor copies, corresponding to all the matricizations of the tensor, can entirely fit in the GPU global memory.

\subsubsection{Canonical Poliyedic Tensor Decomposition (CPD)}
CPD decomposes $\mathcal{X}$ into a sum of single-mode tensors (i.e., arrays), which best approximates $\mathcal{X}$. For example, given 3-mode tensor $\mathcal{X} \in \mathbb{R}^{I_0 \times I_1 \times I_2}$, our goal is to approximate the original tensor as $\mathcal{X} \approx \sum_{r=0}^{R-1} \mathbf{a}_r \circ \mathbf{b}_r \circ \mathbf{c}_r$, where $R$ is a positive integer and $\mathbf{a}_r \in \mathbb{R}^{I_0}$, $\mathbf{b}_r \in \mathbb{R}^{I_1}$, and $\mathbf{c}_r \in \mathbb{R}^{I_2}$.

% \vspace{-5mm}
% \begin{algorithm}
% \DontPrintSemicolon
% Input: A tensor $\mathcal{X} \in \mathbb{R}^{I_0 \times I_1 \times I_2}$, the rank $R \in \mathbb{Z}^{+}$ \;
% Output: CP decomposition $[\![ \mathbf{A}, \mathbf{B}, \mathbf{C} ]\!]$, $\mathbf{A} \in \mathbb{R}^{I_0 \times R}$, $\mathbf{B} \in \mathbb{R}^{I_1 \times R}$, $\mathbf{C} \in \mathbb{R}^{I_2 \times R}$ \;
% \While{\emph{stopping criterion not met}}{
% \textcolor{blue}{// Matricization of $\mathcal{X}$ is different for each factor matrix computation} \;

%     $\mathbf{A} \gets \mathbf{spMTTKRP}(\mathcal{X}_{(0)}, \mathbf{B}, \mathbf{C})$ \;
%     $\mathbf{B} \gets \mathbf{spMTTKRP}(\mathcal{X}_{(1)}, \mathbf{A}, \mathbf{C})$ \;
%     $\mathbf{C} \gets \mathbf{spMTTKRP}(\mathcal{X}_{(2)}, \mathbf{A}, \mathbf{B})$ \;
%    Normalize $\mathbf{A}$, $\mathbf{B}$, $\mathbf{C}$ \;
% }
% % \Return $\mathbf{{A}}$
% \caption{CP-ALS for 3-mode tensors}
% \label{cp-als}
% \end{algorithm}
% \vspace{-4mm}

For each of the three modes, the spMTTKRP operation can be expressed as
% \begin{equation} \label{eqn_spMTTKRP0}
% % \begin{align*} \label{eqn_spMTTKRP}
% \mathbf{\tilde{A}} =  \mathcal{X}_{(0)} ( \mathbf{{B}} \odot  \mathbf{{C}})
% % \end{align*}
% \end{equation}
% \begin{equation} \label{eqn_spMTTKRP1}
% % \begin{align*} \label{eqn_spMTTKRP}
% \mathbf{\tilde{B}} = \mathcal{X}_{(1)} ( \mathbf{{C}} \odot  \mathbf{{A}}), \, 
% % \end{align*}
% \end{equation}
% \begin{equation} \label{eqn_spMTTKRP2}
% % \begin{align*} \label{eqn_spMTTKRP}
% \mathbf{\tilde{C}} = \mathcal{X}_{(2)}  ( \mathbf{{A}} \odot  \mathbf{{B}})
% % \end{align*}
% \end{equation}
\begin{equation} \label{eqn_spMTTKRP0}
% \begin{align*} \label{eqn_spMTTKRP}
\mathbf{\tilde{A}} =  \mathcal{X}_{(0)} ( \mathbf{{B}} \odot  \mathbf{{C}}), \text{ } 
\mathbf{\tilde{B}} = \mathcal{X}_{(1)} ( \mathbf{{C}} \odot  \mathbf{{A}}), \text{ }
\mathbf{\tilde{C}} = \mathcal{X}_{(2)}  ( \mathbf{{A}} \odot  \mathbf{{B}}) \text{ }
\end{equation}

The alternating least squares (ALS) method is used to compute CPD. In a 3-mode tensor, CPD sequentially performs the computations in Equation~\ref{eqn_spMTTKRP0}, iteratively. This can be generalized to higher mode tensors. Note that the matricization of $\mathcal{X}$ is different for each factor matrix computation. In this paper, performing MTTKRP on all the matricizations of an input tensor is called computing MTTKRP along all the modes. The outputs $\mathbf{A}$, $\mathbf{B}$, and $\mathbf{C}$ are the factor matrices that approximate $\mathcal{X}$. $\mathbf{a}_r$, $\mathbf{b}_r$, and $\mathbf{c}_r$ refers to the $r^{\text{th}}$ column of $\mathbf{A}$, $\mathbf{B}$, and $\mathbf{C}$, respectively.

In this paper, we focus on sparse MTTKRP (spMTTKRP), which means the input tensor is sparse and the factor matrices are dense.

\subsubsection{Elementwise computation}\label{background_element_computation}

\begin{wrapfigure}{r}{0.30\textwidth}
\vspace{-6mm}
  \begin{center}
    \includegraphics[width=0.25\textwidth]{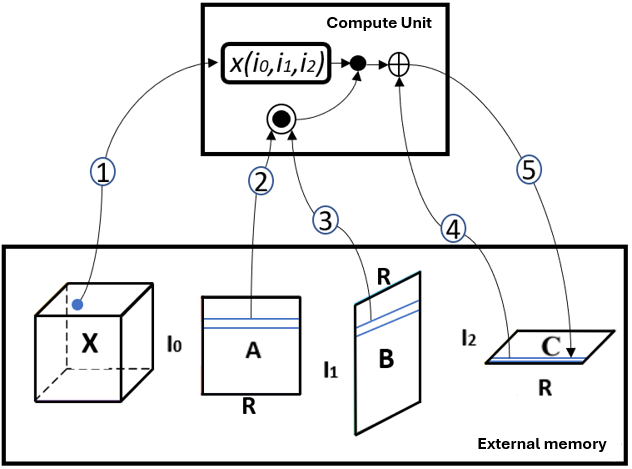}
  \end{center}
  \vspace{-4mm}
  \caption{Elementwise computation~\cite{10.1145/3649153.3649187}}
  \label{element_fig}
  \vspace{-3mm}
\end{wrapfigure}

Figure~\ref{element_fig} summarizes the elementwise computation of a nonzero tensor element in mode 2 of a 3-mode tensor.

In Figure~\ref{element_fig}, the elementwise computation is carried out on a nonzero tensor element, denoted as $\mathcal{X}_{(2)}(i_0,i_1,i_2)$. In sparse tensors, $\mathcal{X}_{(2)}(i_0,i_1,i_2)$ is typically represented in formats such as COOrdinate (COO) format. These formats store the indices ($i_0$, $i_1$, and $i_2$) along with the element value (i.e., $val(\mathcal{X}_{(2)}(i_0,i_1,i_2))$).

To perform the computation, $\mathcal{X}_{(2)}(i_0,i_1,i_2)$ is first loaded onto the processing units (i.e., streaming multiprocessors for GPU) from the external memory (step \circled{1}). The compute device retrieves the rows $\mathbf{{A}}(i_0,:)$, $\mathbf{{B}}(i_1,:)$, and $\mathbf{{C}}(i_2,:)$ from the factor matrices using the index values extracted from $\mathcal{X}_{(2)}(i_0,i_1,i_2)$ (step \circled{2}, step \circled{3}, and step \circled{4}). Then, the compute device performs the following computation:
\[
\mathbf{{C}}(i_2,r) = \mathbf{{C}}(i_2,r) + val(\mathcal{X}_{(2)}(i_0,i_1,i_2)) \cdot \mathbf{{A}}(i_0,r) \cdot \mathbf{{B}}(i_1,r)
\]
Here, $r$ refers to the column index of a factor matrix row ($r < R$). The operation involves performing a Hadamard product between row $\mathbf{{A}}(i_0,:)$ and row $\mathbf{{B}}(i_1,:)$, and then multiplying each element of the resulting product by $val(\mathcal{X}_{(2)}(i_0,i_1,i_2))$. Finally, the updated value is stored in the external memory (step \circled{5}).

% \begin{figure*}
%     \centering
%     \subfigure[Example 3-mode hypergraph]{\includegraphics[width=0.25\textwidth]{images/example_hypergraph.PNG}} 
%     \subfigure[Partitions along Mode $0$]{\includegraphics[width=0.20\textwidth]{images/hyper_mode0.PNG}} 
%     \subfigure[Partitions along Mode $1$]{\includegraphics[width=0.18\textwidth]{images/hyper_mode1.PNG}} 
%     \subfigure[Partitions along Mode $2$]{\includegraphics[width=0.20\textwidth]{images/hyper_mode2.PNG}} 
%     \vspace{-2mm}
%     \caption{Example hypergraph partitioning}
%     \label{example_hypergraph}
%     \vspace{-6mm}
% \end{figure*}

% \begin{figure*}[ht]
% \centering
% % \includegraphics[width=0.9\linewidth]{images/R16_TE.PNG}
% \includegraphics[width=\linewidth]{images/example_hypergraph.PNG}
% \vspace{-5mm}
% \caption{Example hypergraph partitioning}
% \label{example_hypergraph}
% \vspace{-5mm}
% \end{figure*}

% \vspace{-2mm}
\subsection{Related Work}

A. Nguyen et al.~\cite{10.1145/3524059.3532363} propose the Blocked Linearized CoOrdinate (BLCO) format that enables efficient out-of-memory computation of MTTKRP on each mode of the input tensor. In contrast to BLCO, we eliminate intermediate results communication between GPU global memory and GPU SMs and do not require a conflict resolution algorithm like BLCO, which can introduce additional overhead to the execution time.

I. Nisa et al.~\cite{8821030, 10.1145/3295500.3356216} propose a novel tensor format to distribute the workload among GPU threads. Unlike~\cite{8821030, 10.1145/3295500.3356216}, the load balancing scheme proposed in our work reduces the synchronization overhead of GPU SMs and the total GPU global memory accesses.

J. Li et al.~\cite{li2018parti} introduce a GPU implementation employing HiCOO~\cite{8665782} tensor format to accelerate spMTTKRP. HiCOO tensor format uses a block-based tensor format with compression techniques to handle sparse tensors efficiently. Compared with~\cite{li2018parti}, our work reduces the intermediate value communication to the GPU global memory and introduces a novel load-balancing scheme to optimally distribute the total computations among GPU SMs.

 \vspace{-3mm}
\section{Tensor Format}~\label{Data_partitioning}
In this section, we first briefly summarize the notation of the hypergraph representation of a tensor and then use it to describe our load balancing scheme, partitioning, and tensor format.

When performing spMTTKRP on the mode $d$ matricization of the input tensor, we denote the mode $d$ as the output mode and its corresponding factor matrix as the output factor matrix. The rest of the tensor modes are called input modes, and their corresponding factor matrices are called input factor matrices.

\subsection{Hypergraph Representation} \label{sec_hypergraph}

\begin{wrapfigure}{r}{0.25\textwidth}
\vspace{-4mm}
  \begin{center}
    \includegraphics[width=0.25\textwidth]{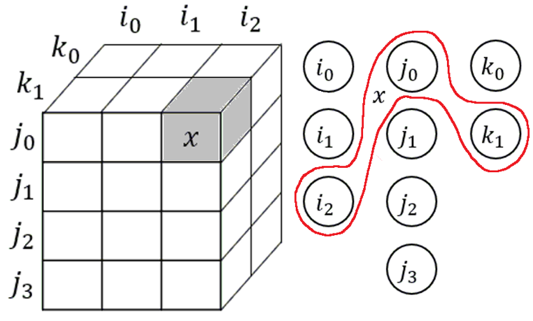}
  \end{center}
  \vspace{-4mm}
  \caption{Example hypergraph of a 3-mode tensor}
  \label{hypergraph}
  \vspace{-5mm}
\end{wrapfigure}

For a $N$ mode tensor $\mathcal{X} \in \mathbb{R}^{I_0 \times \cdots \times I_{N-1}}$, with $|\mathcal{X}|$ nonzero elements,  we consider its corresponding hypergraph, $\mathscr{G}(\textbf{I}, \Upsilon)$ with  vertex set $\textbf{I} = I_0 \cup I_1 \cup \cdots \cup I_{N-1}$ and each nonzero tensor element in $\mathcal{X}$ being represented as a hyperedge in $\Upsilon$. 

Here, $I_d$ is the set of all the indices in mode $d$ and  $|\Upsilon| = |\mathcal{X}|$. Figure~\ref{hypergraph} shows an example representation of a hypergraph for a 3-mode tensor. Note that, in the hypergraph representation, the computation of the output factor matrix of mode $d$ involves performing elementwise operations on all the hyperedges connected to the same vertex in mode $d$ of the tensor.
\vspace{-3mm}
\subsection{Adaptive Load Balancing Scheme}~\label{load_balancing}
In this Section, we propose an adaptive load balancing scheme to optimally distribute the total computations and memory accesses among GPU SMs based on the number of indices in each output mode and nonzero elements of the tensor.

Following the notation introduced in Section~\ref{sec_hypergraph}, consider an input tensor with a corresponding hypergraph of $\mathscr{G}(\textbf{I}, \Upsilon)$ and a target GPU with $\kappa$ number of SMs. In the proposed load balancing scheme, we partition the input tensor into $\kappa$ partitions, where each partition is mapped to a GPU SM.

In our proposed adaptive load balancing scheme, we introduce an adaptive load balancing scheme to distribute the tensor elements depending on the number of indices of the output mode. If the number of indices of output mode is more than or equal to ($I_d \geq \kappa$) the total number of GPU SMs, we distribute the nonzero tensor elements such that the output mode indices are distributed equally among GPU SMs (Load Balancing Scheme 1). Otherwise, if the number of indices of the output mode is less than the total number of GPU SMs ($I_d < \kappa$), the nonzero tensor elements are equally among GPU SMs (Load Balancing Scheme 2). Distributing output indices between GPU SMs without sharing them among SMs avoids global atomic operations between GPU SMs. Meanwhile, output tensor modes with a small number of indices (i.e., $I_d \leq \kappa$) idle GPU SMs during execution time, significantly increasing the total execution time. Hence, even though the equal distribution of nonzero tensor elements among SMs enables global atomic operations, we adapt the load balancing scheme 2 for output tensor modes with a small number of indices.

\subsubsection{Equal Distribution of Indices among Tensor Partitions (Load Balancing Scheme 1)}~\label{adaptive_load_balancing}

In $\mathscr{G}$, for a given mode $d$, the vertices in $I_d$ are ordered based on the number of hyperedges in $\Upsilon$ incident on each vertex. Let us denote the ordered vertex set for mode $d$ as $I_{d-\text{ordered}}$. Subsequently, we iterate through the ordered list, $I_{d-\text{ordered}}$, vertex by vertex, and assign each vertex to a partition in a cyclic fashion. This step effectively partitions the vertices into $\kappa$ index partitions. Next, we collect all the hyperedge incidents on each index partition, which generates tensor partitions. Once the partitioning is complete, we order the hyperedges based on the partition IDs.

\subsubsection{Equal Distribution of Tensor Elements among Tensor Partitions (Load Balancing Scheme 2)}

In $\mathscr{G}$, for a given mode $d$, the hyperedges in $\Upsilon$ are ordered according to the output mode vertex id. Let us denote the ordered hyperedges for mode $d$ as $\Upsilon_{d-\text{ordered}}$. Subsequently, we divide  $\Upsilon_{d-\text{ordered}}$ into $\kappa$ equal size partitions.

Similarly to previous work~\cite{10.1145/3543622.3573179, wijeratne2023dynasor}, we can show that in the proposed load-balancing schemes, the distribution of the nonzero tensor elements is at most 4/3 times the best possible partitioning possible. It also results in the same theoretical tight bound as the theorem in~\cite{graham1969bounds, 10.1145/3543622.3573179}.
\vspace{-3mm}
\subsection{Tensor Representation and Memory Requirements} \label{Tensor_Format_Definition}
We use the COOrdinate (COO) format~\cite{10.1145/3276493} to represent each nonzero tensor element of the input tensor. A tensor of $|\mathcal{X}|$ nonzero tensor elements and $N$ modes is a sequence $x_0, \ldots, x_{|\mathcal{X}|-1}$, where each element $x_i$ is a tuple $\langle p_i, val_i \rangle$. Here, $p_i = (c_0, \ldots, c_{N-1})$ is the indices of the nonzero tensor element in each mode. $val_i$ is the value of the nonzero tensor elements of the tensor at $p_i$. Following COO notation, a single nonzero element requires approximately $|x|_{bits} = \sum_{h=0}^{N-1}\log_2(|c_h|) + \beta_{\text{float}}$ bits, where $\beta_{\text{float}}$ is the number of bits needed to store the floating-point value of the nonzero tensor element. Since each matricization of the tensor requires a unique tensor copy, our work requires $N\times |\mathcal{X}| \times |x|_{bits}$ of GPU global memory to store all the copies of the input tensor.
% \vspace{-5mm}
\section{Parallel Algorithm}\label{subsecparallel_algo}

\subsection{Mode-by-mode spMTTKRP}\vspace{-5mm}
\begin{algorithm}[ht]
    \DontPrintSemicolon
    % \textbf{spMTTKRP}($\mathcal{H}_0$, $\mathcal{B}$, $\textbf{Y}$) \;
    Input: Input tensor copies, $\textbf{T} = \{T_0, T_1 \cdots T_{N-1}\}$ \;
    % \hspace{10mm} Meta data: Tensor Partitions, $\{m_{d}:\forall d\}$\;
    Randomly initialized factor matrices, $\textbf{Y} = \{Y_0, Y_1,...,Y_{N-1}\}$\;
    Output: Updated factor matrices $\hat{\textbf{Y}} = \{\hat{Y}_0, \hat{Y}_1,...,\hat{Y}_{N-1}\}$ \;
    \For{each mode $d = 0, \ldots, N-1$} {
    $T_{in} \leftarrow T_d$ \;
    \For{\text{each partition}, $B_{d,z}$ in $T_{in}$ \textbf{parallel}}{
    \{$Y_d$, $T_{out}$\} $\leftarrow$ \textbf{Thread Block}($B_{d,z}$, $\textbf{Y}$, $T_{out}$) \;
    }
    \_\_Global Barrier\_\_ \;
    }
\caption{Overall Proposed Algorithm}
\label{parallel_alg}
\end{algorithm}
Algorithm~\ref{parallel_alg} shows the parallel algorithm on performing spMTTKRP. Algorithm~\ref{parallel_alg} takes (1) all the input tensor copies $\textbf{T}$ and (2) factor matrices denoted as $\textbf{Y} = \{Y_0, Y_1,..., Y_{N-1}\}$. As shown in Algorithm~\ref{parallel_alg}, the spMTTKRP is performed mode by mode (Algorithm~\ref{parallel_alg}: line 6). In each mode, A thread block (Algorithm~\ref{parallel_alg}: line 7) operates on a tensor partition mapped into a GPU SM during the load-balancing stage~(see Section~\ref{load_balancing}). At the end of all the computations of one mode, the GPU is globally synchronized before the computations of the next mode to maintain the correctness of the program (Algorithm~\ref{parallel_alg}: line 10).
\vspace{-2mm}
\subsection{Mapping Parallel Algorithm to GPU Thread Blocks}~\label{sec_grid_model}~\label{sec_proposed_design}
The basic computing unit of a GPU is a thread. According to the GPU programming model~\cite{cuda2021cuda, ansorge2022programming}, a multi-threaded program is partitioned into blocks of threads (i.e., thread blocks) that operate independently.

\begin{algorithm}[ht]
    \DontPrintSemicolon
\textbf{Thread Block}($B_{d,z}$, $\textbf{Y}$, $T_{out}$):{ \;
    \textbf{Input}: Input tensor partition, $B_{d,z}$\; Factor matrices $\textbf{Y} = \{Y_0, Y_1,...,Y_{N-1}\}$\;

    \textbf{Output}: Updated factor matrix of mode $d$, $Y_d$\;

$nnz = 0$ \;
\For{$nnz < |B_{d,z}|$ \textbf{parallel}}{
    % \If{$a$ of GPU SM is idle}{
    % \textit{Active Block$_a$} $\leftarrow$ $B_{d,z}$ \;
        % \While{nonzero tensor element, $nnz$ < $|B_{d,z}|$}{
    \For{each column, $t$ in thread block \textbf{parallel}}{
        \If{$nnz + t < |B_{d,z}|$}{
        \textbf{Load}($x_i$ at $(nnz+t)$) \;
            $value \leftarrow val_i$ \;
            $p_i = (c_0, \ldots, c_{N-1})$ \;
            \textcolor{blue}{// Elementwise Computation} \;
    \For{input mode $w\in\{0,\ldots,N-1\}\setminus\{d\}$}{

        $vec \leftarrow $ \textbf{Load}(row $c_w$ from $w^\text{th}$ factor matrix) \;
        \textcolor{blue}{// Row $0$ to $R-1$ of the thread block perform independent computations} \;
        \For{each rank $r$ in $R$ \textbf{parallel}}{
            $\ell(r) \leftarrow \ell(r) \times vec(r)$ \;
        }
    }
    }
        \For{each rank $r$ in $R$ \textbf{parallel}}{
        \If{\textcolor{red}{Load Balancing Scheme 1}}{
        $Y_d(c_d, r) \leftarrow \text{Local}\_\text{Update}(Y_d(c_d, r) + \ell(r))$
        }
        \If{\textcolor{red}{Load Balancing Scheme 2}}{
        $Y_d(c_d, r) \leftarrow \text{Global}\_\text{Update}(Y_d(c_d, r) + \ell(r))$
        }
    }
    }
     \textcolor{blue}{// P is the number of columns in a thread block} \;
    $nnz \leftarrow nnz + P$ \;
    }
}

\caption{Parallel Algorithm on the GPU thread block (for mode $d$)}
\label{parallel_alg_grid_op}
\end{algorithm}

In our proposed algorithm, a thread block has a dimension of $R \times P$, where $R$ denotes the rank of the factor matrices and $P$ indicates the number of nonzero tensor elements parallelly loaded to a thread block. Here, each column of the thread block shares the same nonzero tensor element, and each column performs elementwise computation on a nonzero tensor element.

Algorithm~\ref{parallel_alg_grid_op} outlines the computations executed on each GPU thread block. In Algorithm~\ref{parallel_alg_grid_op}, $B_{d,z}$ corresponds to $z^{\text{th}}$ tensor partition in mode $d$. When a GPU SM is idle, a thread block and its corresponding tensor partition are assigned to the SM for computation. Each column in the thread block loads a single nonzero tensor element at a time and shares the nonzero tensor element across the threads in the same column. Each thread in a column extracts the information from the tensor element $x_i$ in COO format (Algorithm~\ref{parallel_alg_grid_op}: line 9-11). Subsequently, each thread block performs elementwise computation (Algorithm~\ref{parallel_alg_grid_op}: line 6-22). To achieve threadwise parallelism, each thread in a column only executes the update operation on a single column of a row of the output factor matrix (Algorithm~\ref{parallel_alg_grid_op}: line 15 - 17). The rows of the input factor matrices are loaded from GPU global memory (Algorithm~\ref{parallel_alg_grid_op}: line 13-14) depending on the indices of the current tensor element ($p_i$) executed in the GPU thread. Each GPU thread block locally updates the output factor matrix (Algorithm~\ref{parallel_alg_grid_op}: lines 15) while maintaining the coherency of each thread block to ensure the correctness of the program. As mentioned in Section~\ref{load_balancing}, equal distribution of output mode indices (Load Balancing Scheme 1) only requires atomic operations inside the same thread block (i.e., Local\_Update); meanwhile, equal distribution of nonzero tensor elements requires global atomic operations (i.e., Global\_Update) among thread blocks. Hence, we use Local\_Update or Global\_Update operations based on the load balancing scheme used in the output mode (Algorithm~\ref{parallel_alg_grid_op}: lines 19-22).
% \vspace{-3mm}
\section{Experimental Results}~\label{experiments}\vspace{-8mm}

\subsection{Experimental Setup}~\label{ex_setup}\vspace{-3mm}

\subsubsection{Platform}\label{sec_platform}
We conduct experiments on the NVIDIA RTX 3090, featuring the Ampere architecture. The platform has 82 Streaming Multiprocessors (SMs) and 10496 cores running at 1.4 GHz, sharing 24 GB of GDDR6X global memory. Table~\ref{table_platforms} shows the details of the platform.

\subsubsection{Implementation}
We develop the source code using the CUDA C++~\cite{cuda2021cuda} and compile it using CUDA 11.8~\cite{fatica2008cuda}.
\begin{wraptable}{r}{0.34\textwidth}
 % \vspace{-3mm}
\caption{Platform specifications}
\vspace{-2mm} 
% \begin{center}
\begin{tabular}{ |c|c|c|c|}
 \hline
 Frequency & 1695 MHz \\ 
  \hline
 Peak Performance &  35.6 TFLOPS \\
 \hline
 On-chip Memory & 6 MB L2 Cache \\
 \hline
 Memory Bandwidth & 936.2 GB/s \\
 \hline
\end{tabular}
\label{table_platforms}
% \end{center}
% \vspace{-3mm}
\end{wraptable}

\subsubsection{Datasets}

We use tensors from the Formidable Repository of Open Sparse Tensors and Tools (FROSTT) dataset~\cite{frosttdataset}. Table~\ref{table3} summarizes the characteristics of the datasets.

 \begin{table}[ht]
 \vspace{-2mm}
\caption{Characteristics of the sparse tensors}
\vspace{-2mm}
\begin{center}
\resizebox{\columnwidth}{!}{
\begingroup
\setlength{\tabcolsep}{6pt} 
\renewcommand{\arraystretch}{1.1} 
\begin{tabular}{ |c|c|c|c|}
 \hline
 \textbf{Tensor} & \textbf{Shape} & \#\textbf{NNZs} \\
 \hline\hline
 Chicago & $6.2K\times 24\times 77\times 32$ & $5.3M$\\ 
 \hline
 Enron & $6.1K\times 5.7K\times 244.3K\times 1.2K$ & $54.2M$ \\ 
 \hline
  Nell-1 & $2.9M\times 2.1M\times 25.5M$ & $143.6M$ \\
 \hline
Nips & $2.5K\times 2.9K\times 14.0K\times 17$ & $3.1M$ \\
 \hline
 Uber & $183\times 24\times 1.1K\times 1.7K$ & $3.3M$ \\
 \hline
 Vast & $165.4K\times 11.4K\times 2\times 100\times 89$ & $26M$ \\
 \hline
\end{tabular}
\endgroup
}
\label{table3}
\end{center}
\vspace{-2mm}
\end{table}

\subsubsection{Baselines}~\label{baselines_exp}
We evaluate the performance of our work by comparing it with the state-of-the-art GPU implementations: BLCO~\cite{10.1145/3524059.3532363}, MM-CSF~\cite{8821030}, and ParTI-GPU~\cite{li2018parti}.

\begin{figure*}[ht]
\centering
\includegraphics[width=\linewidth]{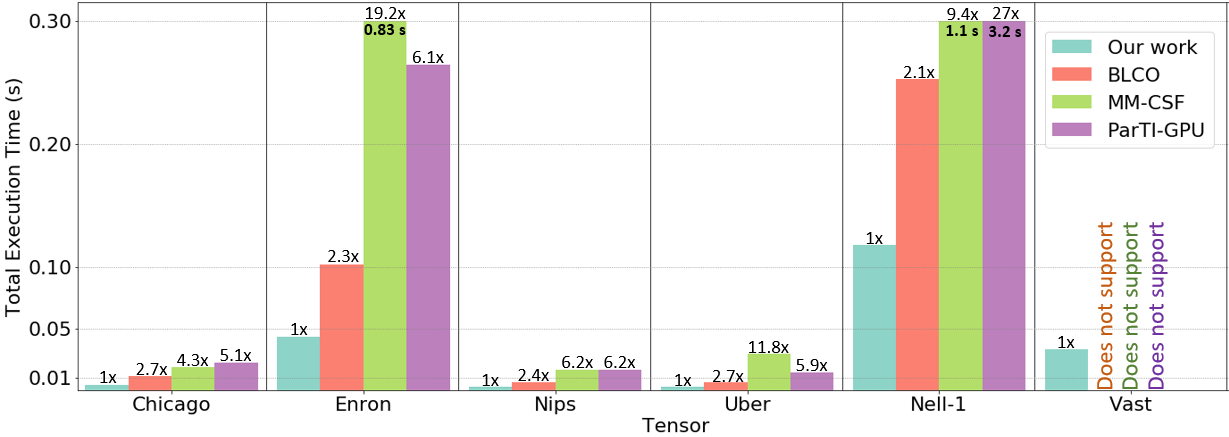}
\vspace{-6mm}
\caption{Total execution time}
\label{R32_TE}
\vspace{-3mm}
\end{figure*}

% \subsubsection{Tensor Partitioning Parameters}
% We set the number of tensor partitions (i.e., $\kappa$) to 82, corresponding to the number of SMs in the device.

% \subsubsection{Tensor Element Representation}
% Following the FLYCOO tensor format representation~\cite{10.1145/3543622.3573179}, we use 32-bit integers to represent the remap ids and the indices. We use 32-bit floating point to represent the value of a nonzero tensor element.

\subsubsection{Default Configuration}~\label{standard_config}
To map the parallel algorithm to the thread block as discussed in Section~\ref{sec_grid_model}, we set $P = 32$, $\kappa = 82$, and $R = 32$ for our selected GPU, RTX 3090.

\subsection{Impact of Adaptive Load Balancing Scheme}~\label{load_balance_results}
\begin{figure}[ht]
\vspace{-3mm}
\centering
\includegraphics[width=\linewidth]{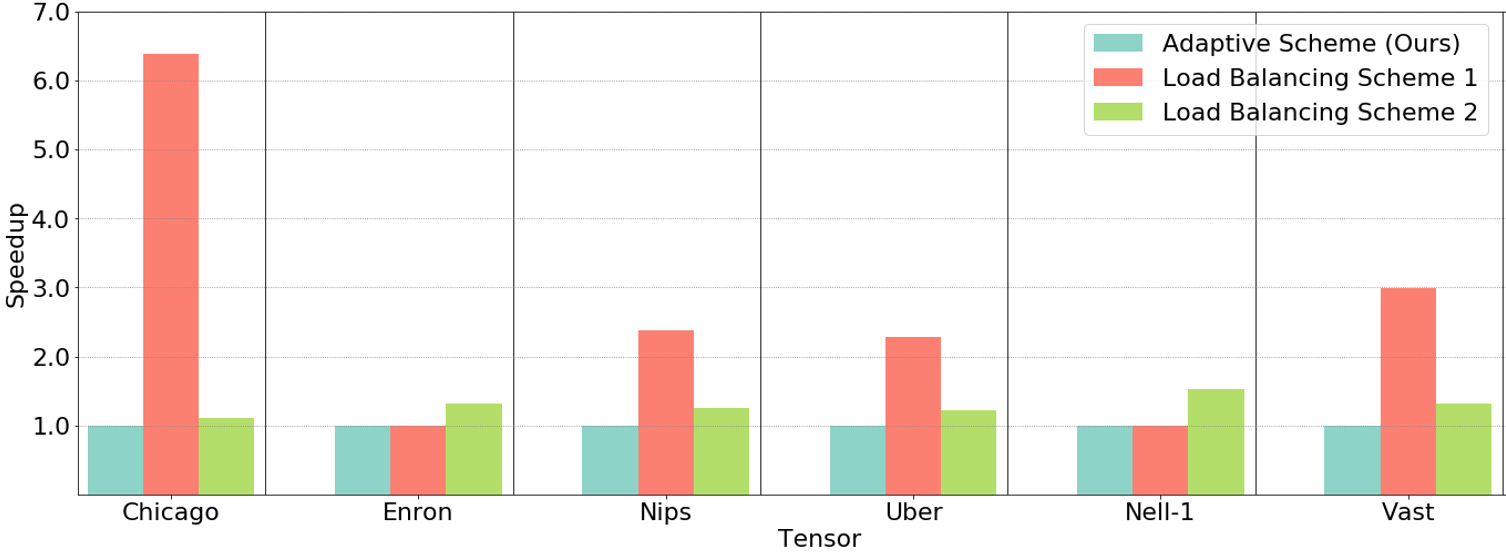}
\vspace{-4mm}
\caption{Impact of the adaptive load balancing scheme}
\label{balancing_scheme}
% \vspace{-2mm}
\end{figure}

In Figure~\ref{balancing_scheme}, we compare our proposed adaptive load balancing scheme (see Section~\ref{load_balancing}) against only using load balancing scheme 1 and only using load balancing scheme 2. Our proposed adaptive load balancing scheme shows a geometric mean of $2.2\times$ and $1.3\times$ speedup in total execution time compared to load balancing schemes 1 and 2, respectively.

Load balancing scheme 1 idles GPU SMs in tensor modes with fewer indices than the number of SMs (in the target GPU) due to the lack of indices to distribute among the GPU SMs. Hence, using only load balancing scheme 1 shows an increased execution time compared with our proposed method for tensors with output modes having fewer indices than the number of SMs (e.g., Chicago, Nips, and Uber). Meanwhile, using only load balancing scheme 2 takes longer than our proposed method for tensors with a large number of indices (number of indices $>$ GPU SMs) due to additional global atomic operations introduced while using only the load balancing scheme 2.

\subsection{GPU Global Memory Requirement}~\label{mem_req}
\begin{figure}[ht]
\centering
\includegraphics[width=\linewidth]{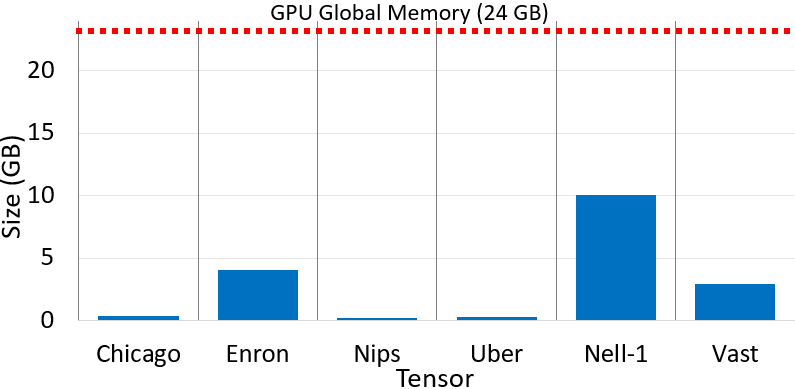}
\vspace{-4mm}
\caption{Total Memory Consumption}
\label{mem_consumption}
\vspace{-6mm}
\end{figure}

Figure~\ref{mem_consumption} shows the total memory required to store all the mode-specific tensor copies and the factor matrices of each input tensor. As illustrated in Figure~\ref{mem_consumption}, all the copies of each input tensor and factor matrices can fit in the GPU global memory of RTX 3090.

% \begin{figure}[ht]
% \vspace{-8mm}
% \centering
% \includegraphics[width=\linewidth]{images/SMutil.PNG}
% \vspace{-5mm}
% \caption{SM throughput comparison}
% \label{smu_fig}
% \vspace{-5mm}
% \end{figure}
% Compute SM throughput is a commonly used metric introduced by NVIDIA Nsight Compute~\cite{iyer2016gpu} for GPU to report the utilization achieved by the SMs while executing a kernel with respect to the theoretical maximum utilization of the selected GPU~\cite{iyer2016gpu}. NVIDIA Nsight Compute provides the achieved throughput of the kernel as a percentage value.

% Figure~\ref{smu_fig} compares the SM throughput of our work for each dataset against the state-of-the-art. We use NVIDIA Nsight Compute to measure the throughput, as mentioned above. In all the datasets, our work shows 1.2$\times$ - 1.4$\times$ and 1.3$\times$ - 2.0$\times$ higher compute throughput than BLCO and MM-CSF, respectively. Our work shows higher throughout due to the minimum SM idle time due to the proposed load balancing scheme and eliminating the intermediate results communication between the SMs. Since the baselines do not support tensors with a large number of modes, we could not report the SM throughput values for BLCO and MM-CSF on Twitch and Vast.
% \vspace{-2mm}

\subsection{Overall Performance}~\label{overall_perf_exp}
In Figure~\ref{R32_TE}, we compare the total execution time of our work with different baselines on RTX 3090. The speedup achieved by our work compared to each baseline is shown at the top of each bar. Similar to the baselines~\cite{10.1145/3524059.3532363, 8821030, li2018parti}, we set the rank of the factor matrices ($R$) to 32. Our work demonstrates a geometric mean of 2.4$\times$, 8.9$\times$, and 7.9$\times$ in speedup compared to BLCO, MM-CSF, and ParTI-GPU. We execute each baseline mode by mode, measure the execution time in each mode, and sum them up to report the total execution time. 

Our work avoids communicating intermediate values among SMs and between SMs and GPU global memory. These intermediate values are stored in the L1 cache and reused with high L1 cache throughput. Also, our load-balancing scheme improves the overall SM throughput, reducing the idle time of the GPU SMs. 

In contrast to BLCO, MM-CSF, and ParTI-GPU, our implementation supports tensors with a number of modes greater than 4.
\vspace{-3mm}
\section{Conclusion and Future Work}~\label{conclusion}
This paper introduced a parallel algorithm for GPUs to accelerate spMTTKRP across all the modes of small tensors. The experimental results demonstrate that Our approach achieves a geometric mean speedup of 2.4$\times$, 7.9$\times$, and 8.9$\times$ in total execution time compared with the state-of-the-art implementations.

Our future work will focus on adapting the proposed parallel algorithm to different computing platforms (e.g., CPU and FPGA). It will ensure that our work can be applied effectively across various hardware.
\vspace{-3mm}
\section*{ACKNOWLEDGEMENT}
This work is supported by the National Science Foundation (NSF) under grant CNS-2009057 and in part by DEVCOM Army Research Laboratory under grant W911NF2220159.

\bibliographystyle{IEEEtran}
\bibliography{reference}

\end{document}